\documentclass[11pt,twoside]{article}

\usepackage{asp2006}
\usepackage{epsf}
\usepackage{psfig}
\usepackage{lscape}

\def\cgcg{CGCG~049-033\,}
\def\ltsim{\raise 2pt \hbox {$<$} \kern-1.1em \lower 4pt \hbox {$\sim$}}
\def\ltapprox{\raise 2pt \hbox {$<$} \kern-1.1em \lower 5pt \hbox {$\approx$}}
\def\gtsim{\raise 2pt \hbox {$>$} \kern-1.1em \lower 4pt \hbox {$\sim$}}
\def\gtapprox{\raise 2pt \hbox {$>$} \kern-1.1em \lower 5pt \hbox {$\approx$}}
\def\arcsec{$^{\prime\prime\,}$}
\def\arcmin{$^{\prime\,}$}

\def\com#1{$^\dagger$}

\begin{document}
\title{ A giant radio jet of very unusual polarization in a
 single-lobed radio galaxy}   
\author{Joydeep Bagchi$^1$, Gopal-Krishna$^2$, Marita Krause$^3$, Chiranjib Konar$^1$, Santosh Joshi$^4$}   
\affil{$^1$IUCAA, Pune University Campus, Pune  411007, India\\ 
$^2$ National Centre for Radio Astrophysics, TIFR, Pune  411007, India\\
$^3$ Max-Planck-Institut f\"{u}r Radioastronomie, Auf dem H\"{u}gel
69, 53121 Bonn, Germany\\
$^4$ Aryabhatta Research Institute of Observational Sciences, NainiTal  263129, India\\ 
}    

\begin{abstract} 
We report the discovery of a very unusual,
predominantly one-sided radio galaxy CGCG049-033.
Its radio jet, the largest detected so far,
emits strongly polarized synchrotron radiation and can be
traced all the way from the galactic nucleus to the hot spot located $\sim 440$
kpc away.  This  jet  emanates from an extremely
massive black-hole $( > 10^9 M_{\odot}$) and forms a strikingly compact
radio lobe. To a surface brightness contrast of at least 20 no radio lobe
is detected on the side of the counter-jet,
which is similar to the main jet in brightness upto the
scale of tens of kpc.
Thus, contrary to the nearly universal trend, the
brightness asymmetry in this radio galaxy
{\it increases} with distance from the nucleus. With several unusual
properties, including a predominantly toroidal magnetic field, this
Fanaroff-Riley type II mega-jet is an exceptionally useful laboratory for
testing the role of magnetic field in jet stabilization and
radio lobe formation.
\end{abstract}

\section{Introduction}
Relativistic jets,  which contain highly collimated streams of plasma
travelling close to the speed of light, are commonly found in  diverse
astrophysical environments. They are  associated with some extreme
relativistic phenomena such as radio galaxies and quasars, microquasars,
pulsars, supernovae and gamma ray bursts.
Bipolar jets in a radio galaxy or quasar are launched from  the
central region of an active galaxy,
probably from a rotating magnetized accretion disk
around a   massive spinning black-hole, the ``central-engine'' \citep{BP82}.
While astrophysical jets have begun to reveal their mysteries, many of
their basic properties, such as their ejection, collimation,
stability  and composition, remain to be understood. Fundamental
questions raised by astrophysical jets include the roles of
magnetic fields in their survival against internal instabilities out to
distances approaching hundreds of kiloparsec from the galactic nucleus
and in the formation of the radio hotspot and lobe due to the jet's
termination.

\section{The remarkable giant radio-jet source CGCG~049-033}
Recently we have reported the discovery of a radio source
associated with the bright elliptical galaxy \cgcg
(at $z = 0.0446$) and displaying a unique combination of
properties, including the largest detected radio jet \citep{Bagchi07}.
This extraordinary object was originally spotted by us
serendipitously in the NVSS database \citep{Condon98}.
Thereafter, we imaged it with the GMRT (Fig.~1, at 1.3~GHz frequency, resolution
3\arcsec to 11\arcsec) and with the Effelsberg radio telescope (Fig.~2,
at 8.4~GHz frequency, resolution 84\arcsec). We also took its
optical spectrum with the 2-meter 
telescope recently set up by IUCAA near Pune (India) at Girawali Observatory. The
massive elliptical galaxy hosting this radio source is
projected 22\arcmin (= 1.1 Mpc) away from the (radio-quiet) central
elliptical galaxy of the rich cluster Abell~2040 at $z = 0.0456$.
The $\sim$8.6\arcmin (= 440 kpc) long radio jet terminates in an
edge-brightened lobe, similar to a Fanaroff-Riley II (FR~II) radio-jet morphology (Fig.~1).

\begin{figure}
\vbox{
\hbox{
\psfig{file=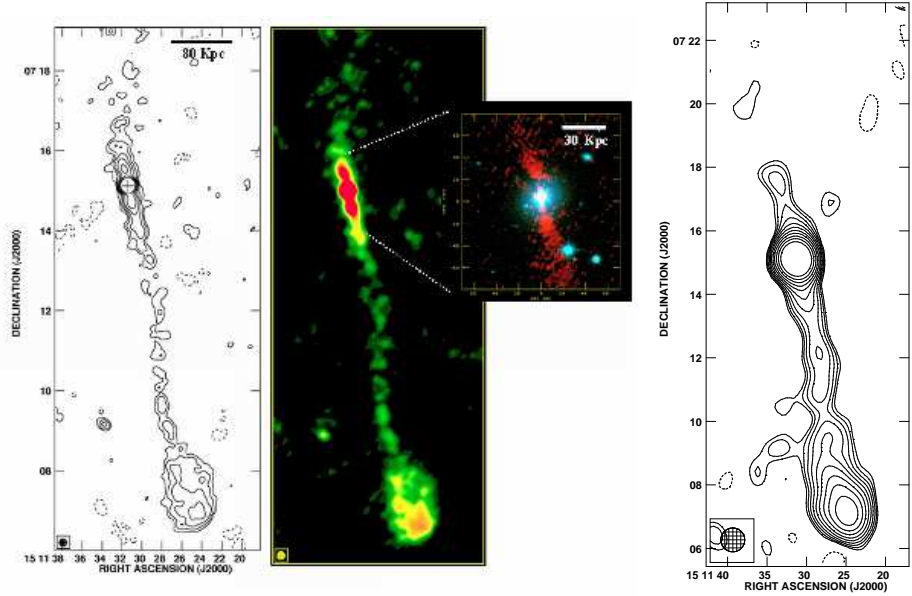,width=0.645\textwidth}
\psfig{file=CGCG049_NVSS_contour.ps,width=0.27\textwidth}
}
}
\caption{GMRT and VLA radio maps of \cgcg. 
{\bf (left):} GMRT 1.28 GHz total intensity map with 
contours at: -0.18, 0.18, 0.36, 0.72,
1.44, 3 and 6 mJy/beam (rms noise: $\sim60$ $\mu$Jy/beam) and the grey 
scale image of \cgcg {\bf (at center)},
both taken with a 11\arcsec\,  beam (FWHM). The inner $\sim$ 2\arcmin\,
region {\bf (inset)} is shown with the 3\arcsec\, resolution
 GMRT 1.28 GHz image overlayed on the  optical r-band SDSS image of host galaxy.
 The `+' marks the radio core position. {\bf(right):} The VLA 1.4 GHz (NVSS) 
 radio map is  shown with contours drawn at: -1, 1, 1.4, 2, 2.8, 4, 5.6, 8 - - - 
  mJy/beam and beam size 45\arcsec\, (FWHM).
}
\label{f1}
\end{figure}

\begin{figure}
\hbox{
\psfig{file=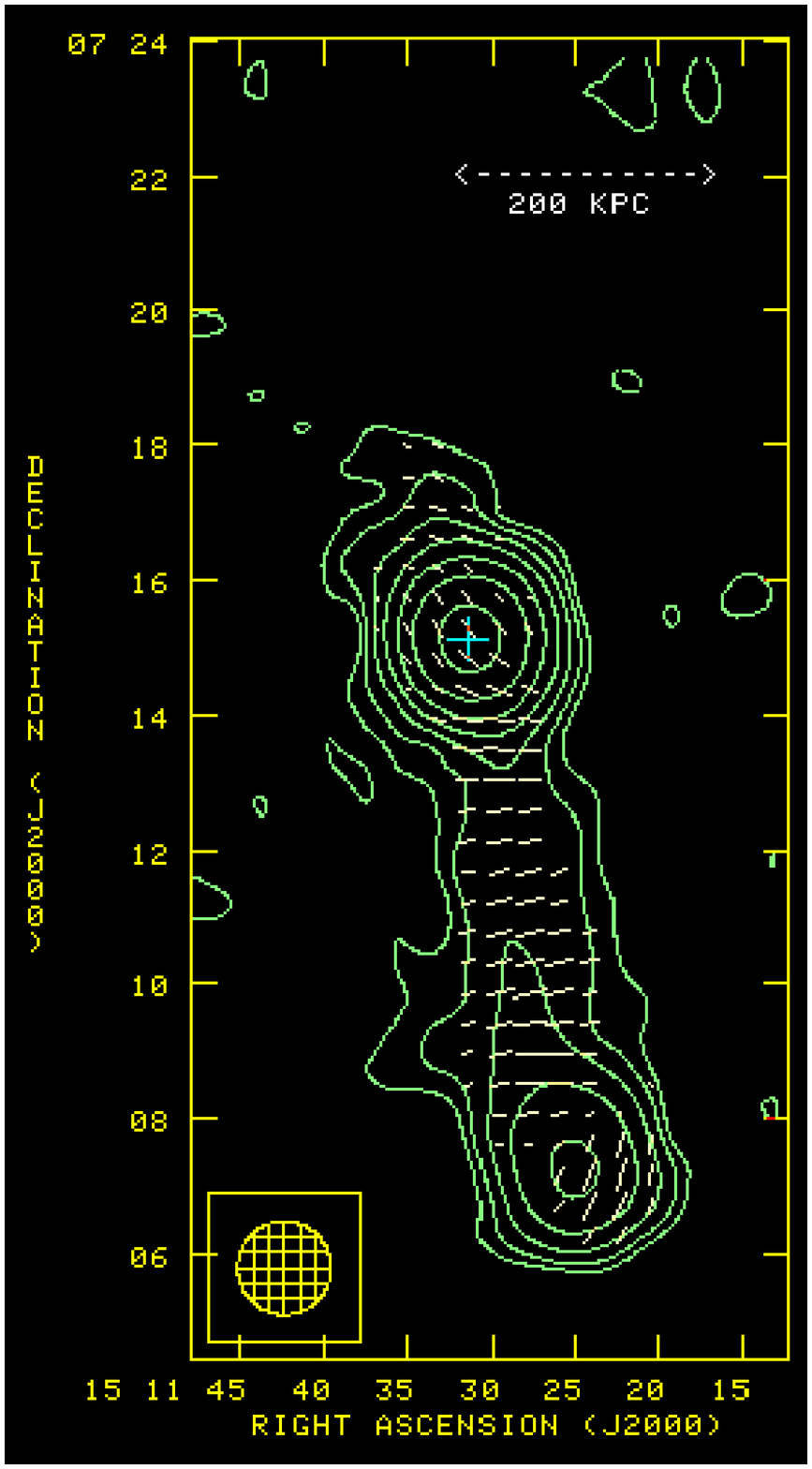,width=0.255\textwidth}
\psfig{file=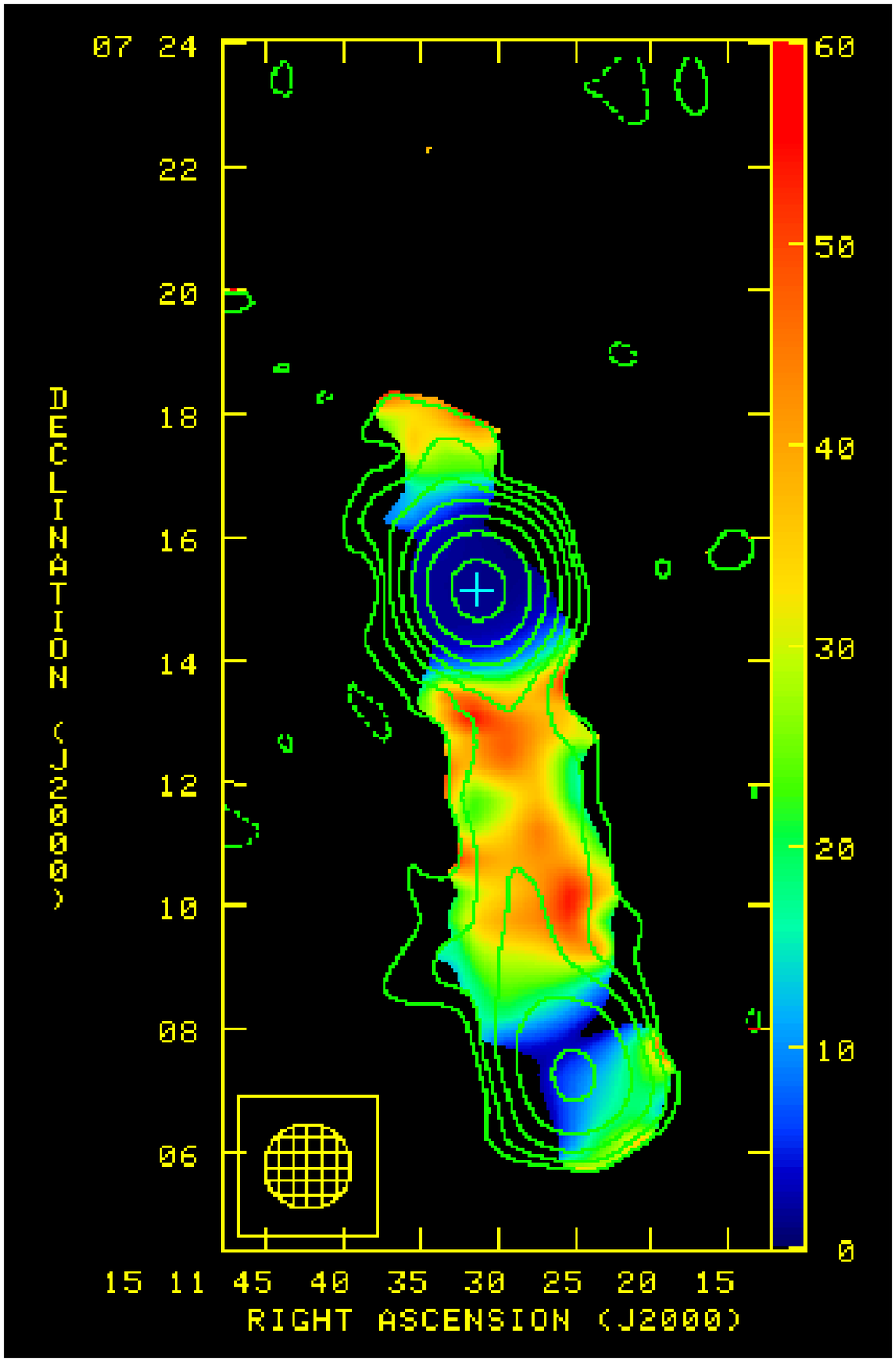,width=0.305\textwidth}
\psfig{file=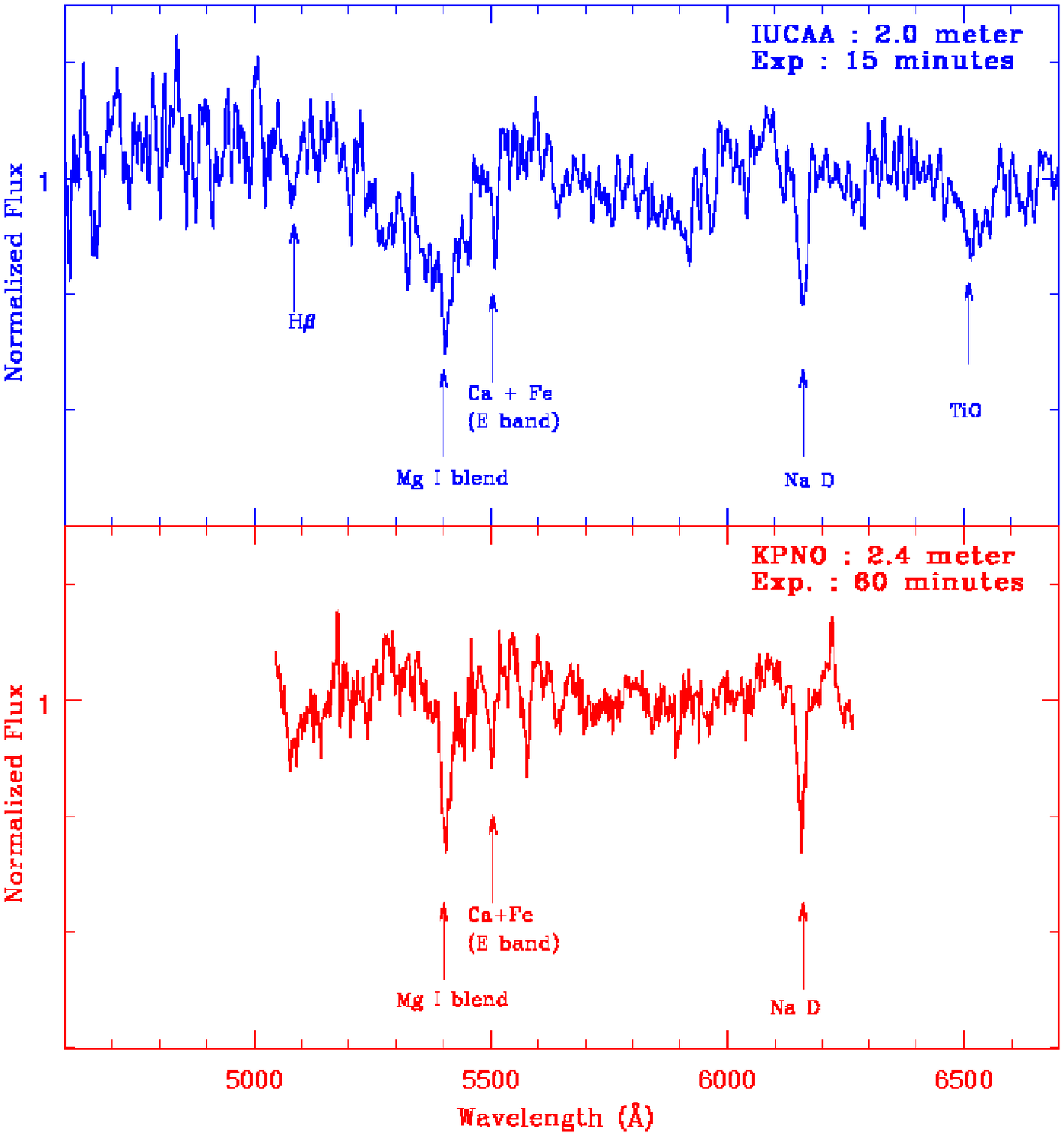,width=0.46\textwidth}
}
\caption{ Effelsberg 8.35~GHz total  power contour map: -0.75, 0.75, 1.5, 3, 6, 12, 24
and 48 mJy/beam, with  {\bf (left):} rotation-measure
corrected magnetic field vectors, having lengths
proportional to the local intensity of polarized flux
(scale: 1\arcsec\, $=$  47.6 $\mu$Jy/beam), and {\bf (center):}
with  image of percentage linear polarization fraction 
(beam: 84\arcsec\, HPBW). {\bf (right):}
Optical spectra of \cgcg taken with IFOSC on the 2-m telescope of IUCAA 
and with the 2.4-m Hiltner/KPNO telescope (see \citet{Bagchi07}). 
}
\label{f2}
\end{figure}

\section{Why the radio source associated with
galaxy CGCG 049-033 is extraordinary?}

Below we highlight a few  striking properties of \cgcg\,
which, when taken together, set it apart from nearly all the known FR~II
radio galaxies:

$\bullet$ Its remarkably well collimated radio jet is not only the largest detected jet, it is
also very strongly polarized ($p\sim$20 to 50\% at 8 GHz, inspite of
the  averaging by the large 84\arcsec\, beam). Furthermore,
The observed strong linear polarization and its orientation imply a
well organized transverse (toroidal) magnetic field for this
giant radio jet. The  projected magnetic field orientation is predominantly
orthogonal to the jet (Fig.~\ref{f2}),
contrary to the norm for FR~II jets \citep{Bridle94}.
Thus, the jet appears to be a truly rare example where an
extremely well ordered toroidal/helical magnetic field
configuration is able to persist out to  a scale $>$100~kpc  
from the nucleus.

$\bullet$ The radio lobe inflated by this giant jet is strikingly 
compact with hardly any back-flow.
This could be result of ``blocking" of the back-flow of its
synchrotron plasma by a toroidal magnetic field \citep{NK97}, as
hinted to by the polarization map (Fig.~\ref{f2}).

$\bullet$ The two radio jets appear quite similar upto the initial
 $\sim$100 kpc, but thereafter the northern jet undergoes an
abrupt fading. Any lobe formed by it could not be detected
even in the Effelsberg single-dish map, as well as  GMRT
and NVSS maps (Fig.~1 and 2). This implies an
extreme factor ($>20$) for the lobe flux asymmetry. Thus,
 contrary to the nearly universal trend, the brightness
 contrast between the two opposite sides of this double
radio source {\it increases} with distance from the nucleus.

$\bullet$ The optical spectrum reveals that the host 
galaxy harbours a supermassive nuclear 
black hole of
mass $\sim 2 \times 10^9$ M$_\odot$.
This is reflected in the  very large stellar velocity dispersion 
$\sigma_{\ast}$ = 375$\pm$35 km s$^{-1}$ of absorption lines and 
large bulge luminosity in K-band ($M_{K}=-26.22\pm0.044$, 
$L_{K,bulge}= 6.5 \times 10^{11} L_{K,\odot}$).

$\bullet$ The observed core-to-lobe flux density ratio, $f_c = 1.67$ at 1.28 GHz is
an extreme value; for FR~II radio galaxies with matching radio lobe power to
CGCG~049-033, $f_c$ is typically only $\sim 0.1$ \citep{Zirbel_Baum95}.
The large excess in $f_c$, by a factor
$F \sim 17$, could be attributed to relativistic beaming which can boost
the core flux by a factor $\sim \delta^n$ (usually {\it n}$~\sim 2$ for
compact flat-spectrum jets, and the Doppler factor is
$\delta = [\Gamma_{j} (1-\beta_j \, 
cos \theta)]^{-1}$, where  $v_{j}= c \beta_j$ is the bulk speed of jet
and $\Gamma_{j}$, the bulk Lorentz-factor). The corresponding
viewing angle $\theta$ of the nuclear jet
\citep{Giovannini_1994}: cos ($\theta$) = $[0.5 + (\sqrt{F} -1)/\beta_j]/\sqrt{F}$,
is then quite small ($\theta$ \ltsim 28$^{\circ}$, for $\beta_j$ \ltsim 1).
Such small angles  are jointly disfavored by  the projected
giant radio size, the apparent symmetry of the inner jets and the
identification with a galaxy \citep{Barthel_1989}. Jet bending on
parsec scale can result in a strongly  Doppler boosted
nuclear jet (despite the large projected size of source). However,
the observed straightness and good alignment
of the inner kiloparsec jets disfavor this scenario (Fig.~\ref{f1}).


\section{Outlook for future}
A higher resolution mapping of the
polarization structure of the jets and the radio lobe 
will  reveal what role a toroidal/helical magnetic field might
play in stabilizing this remarkable radio jet upto
an unprecedented scale of $\sim$400 kpc.
The observed quasi-regularly spaced ($\sim 40$ kpc) sequence of radio knots
(Fig. 1) as well as its excellent
collimation make this mega-jet
a prime target for probing jet confinement mechanisms.
A deeper radio imaging of this jet could be used to test viability of
the theory of jet stabilization upto  $>$100 kpc scales
via a spine-sheath type flow \citep{Hardee2007}, or  the jet collimation
by a surrounding high pressure ambient medium, leading to a  sequence of reconfinement
shocks \citep{Komissarov_Falle_1998}.
An X-ray imaging study with {\it Chandra} of the inner $\sim$100 kpc scale
 jet and counter-jet region, the associated AGN and the hot gaseous 
 halo of the  host galaxy is being planned.
\acknowledgements 
We thank the operations teams of the GMRT (NCRA -- TIFR), Effelsberg (MPIfR) 
and IUCAA Girawali 
observatory (IGO) for help during observations. 


\end{document}